# Abdus Salam: A Reappraisal
## PART II
## Salam's Part in the Pakistani Nuclear Weapon Programme

Norman Dombey[+]

Physics and Astronomy Department

University of Sussex

Brighton BN1 9QH

December 10 2011

## ABSTRACT

Salam's biographies claim that he was opposed to Pakistan's nuclear weapon programme. This is somewhat strange given that he was the senior Science Advisor to the Pakistan government for at least some of the period between 1972 when the programme was initiated and 1998 when a successful nuclear weapon test was carried out. I look at the evidence for his participation in the programme.

---

[+] Email  n.dombey@sussex.ac.uk

Pakistan possesses nuclear weapons. It conducted a series of test explosions in May 1998 in the Ras Koh Hills in Chagai, Balochistan and is believed to have had operational nuclear weapons since March 11 1983, when it successfully carried out a cold test[1, 2] of a weapon design. What part, if any, did Salam play in this?

On January 20 1972 immediately after the Indo-Pakistan war which led to the secession of Bangladesh, Zulfikar Ali Bhutto, the new President, called Pakistan's senior scientists to a conference at Multan[3] and asked them to begin work on a weapon. Munir Khan[4] had told Bhutto back in 1965 that IAEA's inspections of Indian nuclear facilities could only be understood in terms of an Indian weapon programme but Ayub Khan had refused to follow suit. Salam was at Multan. The then Chairman of the Pakistani Atomic Energy Commission Ishrat Hussain Usmani refused at Multan to have anything to do with nuclear weapons: he asked how could Pakistan make a nuclear weapon when it had no industrial infrastructure. Bhutto immediately replaced him as Chairman of PAEC with Munir Khan who was told to report directly to Bhutto.

When Salam was first appointed by Ayub Khan as Presidential Science Advisor, he and Usmani had worked closely together to establish civil nuclear power in Pakistan and had set up the Pakistan Institute of Nuclear Science and Technology (PINSTECH) at Nilore, near Islamabad, whose first reactor went critical in 1965. They had also collaborated on a scholarship scheme which sent Pakistani nuclear students to western countries to study advanced topics. So Salam found himself at the Multan meeting with two close associates, one of whom was totally opposed to Pakistan using its scarce resources on a weapon programme while Munir Khan, who had been a friend of Salam since the 1940s, was leading the programme.

Which way did Salam jump? The official story given in his biography by Gordon Fraser[5] is that he sided with Usmani[6]. Similarly, the biography by Jagjit Singh[7] says that "Both Usmani and Salam disagreed with this [nuclear weapon] policy which they felt was a flagrant misuse of atomic energy"[8] That is what Salam himself would say if asked. Moreover he and IAEA Director Sigvard Eklund had been awarded the Atoms for Peace Medal in 1968; he had been awarded the Peace Medal of Charles University, Prague in 1981, and he had met his second wife Louise Johnson at an anti-nuclear proliferation meeting in London in 1962.

Yet another version is given by Weissman and Krosney[9], where it is claimed that Bhutto sent an emissary to try to convince Salam that the Multan meeting was a pretence and that it was really peaceful in intent.

---

[1] R M S Azam, *When Mountains Move – The Story of Chagai*, Pakistan Defence Journal June 2000 http://www.defencejournal.com/2000/june/chagai.htm
[2] For a discussion of cold testing see N Dombey and O Rabinowitz, *Testing Times* The World Today, February 2011, p.27
[3] Gordon Correra, *Shopping for Bombs* Hurst & Co London , 2006, p.9
[4] For Salam's relationship with Munir Khan see Part I, p.8
[5] Gordon Fraser *Cosmic Anger Abdus Salam –The first Muslim Nobel Scientist* Oxford 2008
[6] Fraser (ibid), p.250
[7] Jagjit Singh, *Abdus Salam A Biography* Penguin Books India 1992
[8] Singh (ibid) p. 36
[9] Steve Weissman and Herbert Krosney, *The Islamic Bomb* Times Books New York,1981, p.46

Both stories are inherently implausible. As Presidential Science Advisor[10] Salam naturally played a central role in Bhutto's principal scientific project after January 1972, namely the nuclear weapon programme. Furthermore there is substantial evidence to back this up.
In 1980 the American columnist Jack Anderson published an article[11] in the Washington Post (which was published in Europe in the International Herald Tribune) entitled 'Pakistan Near Entry in the Atomic Club'. He had very good contacts in the US government and quoted US intelligence sources drawing attention to A Q Khan's theft of centrifuge blueprints from Almelo in Holland. Anderson wrote 'Under Khan's guidance, and with the help of 1979 Nobel Prize-winning physicist Dr Abdus Salam, the Pakistanis are so far along the nuclear trail that military-scientific teams have already been looking for suitable desert expanses for an underground test explosion.' At the time very few knew about A Q Khan. Anderson pointed to Libya's help for the project and that it was likely that Pakistan would share its nuclear capability with Libya: this turned out to be the case.

I happened to be visiting ICTP when Anderson's article was published and I drew Salam's attention to the article. He immediately wrote a letter to Anderson with a copy to me denying any knowledge of A Q Khan's activities and saying that he had never even met him. At least the latter was probably true since Salam's contacts were with Munir Khan and the PAEC, not with the enrichment plant at Kahuta where A Q Khan was based. As far as I know Salam's letter was not published either by the Washington Post or the International Herald Tribune. Anderson (together with Jan Moller) subsequently repeated the claim that Salam had helped develop the bomb[12] after the tests. Anderson and his CIA-informants clearly had not taken Salam's denial of involvement very seriously

Evidence that Salam approved of the Pakistani weapon programme [if not that he directly participated in it] comes from the Iraqi scientist Khidir Hamza. Hamza was on the CIA payroll[13] for a time so extracts from his his book[14] have to be double checked but neither he nor the CIA seems to have any reason to fabricate his encounters with Salam. Hamza writes that Salam had worried about Jafar, the head of Iraq's enrichment programme, who was imprisoned by Saddam for 20 months. That was true: Jafar had worked as a post-doctoral research fellow in particle physics at Imperial College where he would have met Salam, from July 1970 until he returned to Baghdad in 1975[15]. Hamza also writes that Salam said on a visit to Baghdad 'Why don't you guys finish the job and make an atomic bomb….Israel has a few at the very least. You have not only the right but the duty to defend your country'[16]. That was indeed the logic of the Pakistani programme with the substitution of India for Israel'. Salam had visited Baghdad in April 1975 to attend a conference on the peaceful uses of nuclear energy: just because he had fallen out with Bhutto the previous year about the treatment of Ahmadis doesn't mean that he disagreed with Bhutto's nuclear strategy. Hamza also records that Salam told a story about a visit to China just after the first Chinese H-bomb test[17]. Salam did visit China in summer 1966 en route to the Conference on High Energy

---

[10] In 1973 the Pakistani constitution was changed from a presidential system to a parliamentary system: Bhutto became Prime Minister while Salam became Governmental Science Advisor.
[11] Jack Anderson, Washington Post, April 11 1980 p.B9
[12] Jack Anderson and Jan Moller, Deseret News, June 28 1998
[13] N. Dombey, *Saddam's Nuclear Incapacity*, London Review of Books 17 September 2002
[14] Khidir Hamza, *Saddam's Bombmaker* TOUCHSTONE New York 2001
[15] Profile: Jafar Dhia Jafar, *Science* **309** 2158 (2005)
[16] Hamza *op. cit.* p.104
[17] Hamza *op. cit.* p104

Physics in Berkeley, California in August/September that year. China first tested a weapon with a thermonuclear component in May 1966[18] so that checks.

There is now more information available about who did what in the Pakistani programme. The well-connected journalist Shahid-ur-Rehman wrote 'Long Road to Chagai'[19] which was based on interviews with most of the key players, including A Q Khan: there is also the detailed article by Rai Muhammed Saleh Azam[1] in the Pakistan Defence Journal which is probably the PAEC response to the way A Q Khan was claiming much of the credit for himself and Kahuta. This gives the PAEC view about who did what. The two accounts agree in their description of Salam's central role in the project in the early stages of the programme.

In both accounts, Hafeez Qureshi, the head of the Radiation and Isotope Applications division of PINSTEC was summoned by Munir Khan in March 1974 and told that 'he had been picked up to start work on a project of national importance'. His office would be located at Wah near Rawalpindi, a site chosen since 'you would need a lot of explosives': Wah was where the Pakistan Ordnance Factories was based. Salam was present at the meeting, as was Riazuddin, then a senior official in the Pakistan Atomic Energy Commission. The work at Wah came to be known as the work of 'the Wah Group' and 'started by carrying out research and development of the explosive for use in the nuclear device'. Qureshi went on to head the teams who carried out cold tests between 1983 and 1990. He was interviewed by Shahid-ur-Rehman.

The counterpart to Qureshi's explosives group was the Theoretical Group under Riazuddin, described by Shahid-ur-Rehman as 'Pakistan's greatest living scientist' [this is after Salam's death] who 'prepared the first design of a bomb'.

Shahid-ur-Rehman interviewed Riazuddin several times and while he got somewhat confused about the technical terms used, tells a consistent story. He reports that a few months after Multan, Salam visited Pakistan for a meeting with Bhutto and Munir Khan. Then in October 1972 Salam summoned Riazuddin and Masud Ahmad, who were both working at ICTP, to his office--at the time Riazuddin was a Senior Associate and Masud was a Fellow—where he 'informed them about the Pakistan government's political decision to start working for the nuclear option'. In particular 'in the tradition of the Manhattan Project, a Theoretical Group would be set up to carry out R and D for the bomb project and Dr Riazuddin would head it.' On his return to Pakistan Riazuddin met Munir Khan for a briefing 'and plunged into the R and D of designing the bomb while continuing in his position at the Quid-e-Azam University…Masud was directed by Dr Salam to return to Pakistan and join PAEC. He returned, loaded with books on the Manhattan Project provided by Dr Salam and was posted at PINSTECH.'

Shahid-ur-Rehman asked Riazuddin to specify the work of the Theoretical Group. Riazuddin replied 'We were the designers of the bomb, like the tailor who tells you how much of the material is required to stitch a suit. We had to identify the fissile material; whether to use plutonium or enriched uranium; which method of detonation; which explosive; what type of tamper and [explosive] lens to use; how material will be compressed; how shock waves will be created; what would be the yield".

---

[18] R S Norrris, A S Burrows and R W Fieldhouse, *British, French and Chinese Nuclear Weapons*, Vol. V. Westview San Francisco 1984, p. 420

[19] Shahid-ur-Rehman, *Long Road to Chagai* Shahid-ur-Rehman (Islamabad) 1999

"The work of the theoretician is to achieve the maximum yield with the minimum critical mass; to identify the scarce materials [needed]. For example, the critical mass can be greatly reduced by using beryllium as a tamper since it is a very good neutron reflector…..However, beryllium reflectors are difficult to make because it is very brittle and toxic and difficult to handle. Thus the first design prepared by Theoretical Group used U-238 as reflector".

According to Riazuddin an early decision was taken to use implosion to detonate the weapon, rather than using the simpler gun method which was less economical in fissile material. The decision was taken in December 1973 when Salam took a copy of the New Yorker article 'The Curve of Binding Energy'[20] with him to Pakistan. The article was based on interviews with the Los Alamos weapon designer Ted Taylor: in it Taylor discusses the merits of different fissile materials and modes of detonation.

So it seems that Riazuddin was the principal designer of the implosion mechanism for the Pakistani weapon. Riazuddin had spent his whole career working with Salam: first at Punjab University as an undergraduate; then he obtained an M.Sc. in Applied Mathematics at the Punjab University and a Ph. D. in Theoretical Physics from Cambridge, both under Salam's supervision, and finally at ICTP. Both Riazuddin and Munir Khan spoke at Salam's memorial meeting at ICTP after his death. It is not likely that such close colleagues and friends would disagree about such a fundamental issue as the weapon programme without affecting their friendship.

Salam did break with Bhutto in September 1974 when a law was passed declaring that Ahmadis were not Moslems. In response Salam resigned most of his governmental posts. But from the weapon programme's inception in January 1972 until at least September 1974 and possibly beyond[21] it seems that he not only accepted its logic that Pakistan's nuclear weapons were necessary given that India had overwhelming conventional capability but actively helped to achieve the programme's goal.

Given Salam's use of ICTP's resources to further his prospects of a Nobel Prize, this raises the question of whether Salam, Riazuddin and Masud Ahmad, and possibly others, were using ICTP resources to develop Pakistani nuclear weapons. If this were the case, it would have been in violation of IAEA's fundamental objective[22] that "It shall ensure, so far as it is able, that assistance provided by it or at its request or under its supervision or control is not used in such a way as to further any military purpose". The IAEA's main function[23] is "To encourage and assist research on, and development and practical application of, atomic energy for peaceful uses throughout the world".

The question of whether ICTP's resources could be used for nuclear weapon purposes was raised by Steve Coll in the Washington Post on December 24 1992[24]. He asked "whether some of these Third World government scientists, in addition to peaceful research, are carrying out in Trieste work related to nuclear weapons, missile systems or other military technologies? Salam was strangely noncommittal in his reply. He did not try to deny that this was possible. Salam said that his research center follows a "policy of ignoring" whether visiting Third World nuclear and other scientists are working on civilian or military projects. "We have this official policy that work must be done for peaceful purposes, but it's more official than kept up because it's difficult to keep up," Salam said. This is because there is no practical way to distinguish between military or peaceful purposes in the kind of sophisticated nuclear physics and science research that the Trieste center sponsors". This statement seems to me to be dissembling. Compare this with the statement by his successor when the same question was raised:[25] "ICTP does not work, and has never worked, on nuclear technology. ICTP categorically denounces all destructive use of science, nuclear and otherwise, and plays no role in anyone learning anything specific about any weapons programs. All its nuclear-related programs are held in cooperation with IAEA, the world's watchdog agency in charge of preventing the spread of nuclear weapons; without exception, these activities are on peaceful applications in agriculture, climate, energy, medicine and water. The decision on what is an allowable program is made within the guidelines of IAEA, and the selection of participants is done by IAEA subject to its rules".

---

[20] John McPhee, Profiles, "I-THE CURVE OF BINDING ENERGY," *The New Yorker*, December 3, 1973, p. 54; "II- December 10, 1973, p. 50 III-;December 17, 1973, p. 60
[21] R S N Singh claims that Salam initiated cooperation between China and Pakistan on weapon-related issues on a visit to China in 1978
http://www.indiandefencereview.com/geopolitics/Pakistans-Nuclear-and-Missile-Weapons-Programme.html
[22] IAEA Statute, II
[23] IAEA Statute, IIIA
[24] Steve Coll, The Washington Post, December 24 1992, p. A08

The question which needs answering is not whether ICTP had a programme of nuclear technology for civilian purposes which could also be used for weapon purposes. A screwdriver can be used for military purposes but that does not make it a weapon. The question is whether Salam, Riazuddin and Masud Ahmad used ICTP's resources in support of the Pakistani nuclear weapon programme.

At Los Alamos National Laboratory in the US and at the Commisariat d'Atomique Energie's laboratory at Saclay nuclear weapon programmes and civil nuclear programmes co- exist side by side. But everyone participating in them knows on which he or she is working. Funding is clearly delineated. The record of how Salam won his Nobel Prize does not suggest that he was particularly scrupulous in determining whether a particular activity was undertaken for the benefit of science in developing countries or to advance his own research in elementary particle theory. What about his activities as Science Advisor to Pakistan's Government? A Pakistani physicist who was close to Salam, Riazuddin and their colleagues told me[26] that "pre-1974, Salam was anti-Hindu, anti-India, pro-Pakistan, pro-Bomb. He was present at the 1972 meeting with Bhutto in Multan, then brought together a bunch of good Pakistani physicists and made them into a theoretical task force that looked into various physics issues dealing with the bomb, and used ICTP as a platform for this. I know that Salam never did any Bomb-related calculations, but he did attend meetings of the group and asked questions. Post-1974 he changed fundamentally, gradually becoming critical of defence funding and also of the Bomb. He also felt friendlier towards India. He lost his clout almost entirely by the end of the 1970's".

That's likely to be an accurate picture. It leaves open the extent to which ICTP was used for the Pakistani weapon programme and if it were, who knew.

Salam's main legacy is ICTP and the support it gives and has given to young theoretical physicists from developing countries for fifty years. It seems to me that if ICTP had indeed been used as an adjunct of Pakistan's nuclear programme, it is important to take appropriate steps to ensure that this situation could not recur with any other developing country.

---

[25] K.R. Sreenivasan, ICTP Press Release 27/07/2005
[26] email to N Dombey, 26 June 2011